\documentclass[twoside,reqno]{qts-proc}
\usepackage{epsfig,cite}
\usepackage{amssymb,amsmath}
\usepackage{times}
\begin{document}
\sloppy \raggedbottom
\setcounter{page}{1}

\newpage
\setcounter{figure}{0}
\setcounter{equation}{0}
\setcounter{footnote}{0}
\setcounter{table}{0}
\setcounter{section}{0}



\title{Boundary state in open string channel and 
open/closed string field theory}

\runningheads{H. Isono and Y. Matsuo}{Open boundary state
and string field theory}

\begin{start}


\coauthor{Hiroshi Isono}{1},
\author{Yutaka Matsuo}{1}

\address{Department of Physics, Faculty of Science, University of Tokyo,
Hongo 7-3-1, Bunkyo-ku, Tokyo 113-0033, Japan}{1}


\begin{Abstract}
We generalize the idea of boundary states to open string
channel.  They describe the emission and absorption 
of the open string in the presence of intersecting D-branes.
We study the algebra between such states under
the star products of string field theory and confirm
that they are projectors in a generalized sense.
Based on this observation, we propose a modular dual description
of Witten's open string field theory which seems to be
an appropriate set-up to study D-branes by string field theory.
\end{Abstract}
\end{start}


\section{Introduction}

It is almost needless to emphasize the importance
of D-branes in understanding the nonperturbative dynamics
in string theory.  They are essential to the study of 
the strong gravity region where the coupling constant
becomes strong, for example, near the horizon of black holes
or the vicinity of time-like big-bang singularity.

D-branes are also useful since we can treat them 
exactly from 2D conformal field theory.
They are described by boundary states $|B^c\rangle$ 
in the closed string channel\cite{r:BS}. They are
characterized by,
\begin{equation}\label{bc1}
T_{\sigma\tau}|B^c\rangle=0
\end{equation}
or in the Fourier modes of Virasoro algebra,
\begin{equation}\label{bc2}
 (L_n-\tilde L_{-n})|B^c\rangle =0\,.
\end{equation}
The suffix $c$ is attached to indicate that it belongs to
closed string Hilbert space.
There are a large amount of references where
various properties of such states were studied.

As far as we know, the boundary state condition 
(\ref{bc1}) is considered only in the closed string channel.
It is natural in a sense that D-brane plays as a source of
emission/absorption of closed string and the boundary state
describes such a process.  For example, an inner product
between boundary states, $\langle B^c|q^{L_0+\tilde L_0}
|B^c\rangle$, describes closed string propagation between
D-branes. As we shall see below, however,
it is actually possible to consider a similar process
for the open strings.

We consider the situation where
two D-branes $\Lambda$ and $\Sigma$
intersect each other.
\begin{figure}[b]
\centerline{\epsfig{file=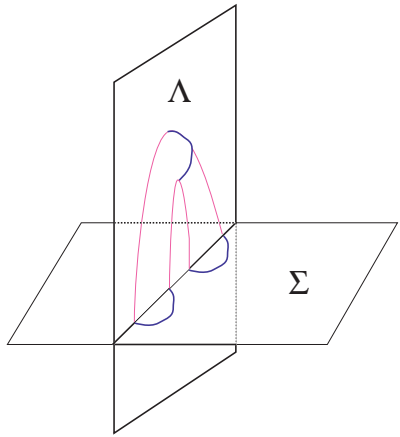,scale=0.3}}
\caption{An open string on D-brane $\Lambda$ is created/absorbed 
on another D-brane $\Sigma$}
\label{fig:open_BS}
\end{figure}
One may consider a physical process (1)
an open string on D-brane $\Lambda$ is emitted from D-brane $\Sigma$,
(2) it propagates on the world volume of  $\Lambda$ and, 
(3) it is  absorbed on $\Sigma$ (fig.\ref{fig:open_BS}).
As the closed string amplitude,
such a process would be described by an inner product of
boundary states as,
\begin{equation}
 {}_{\Lambda\Lambda}^{(\Sigma)}\langle B^o| q^{L_0^{(
\Lambda\Lambda)}}|B^o\rangle^{(\Sigma)}_{\Lambda\Lambda}\,,
\end{equation}
where $|B^o\rangle^{\Sigma}_{\Lambda\Lambda}$ 
belongs to the open string Hilbert space
with both ends at D-brane $\Lambda$.
We will call such states as {\em open boundary states} or OBS
in short. The first purpose of this paper is to define
such states and give basic examples.

One of our motivations to introduce such states is
the description of the D-branes from string field theory.
From some years ago, there have been considerable efforts in
this direction.  A proposal which was examined carefully was vacuum
string field theory (VSFT)\cite{r:VSFT}, where D-branes are described as
the projectors of Witten's star product.  Careful studies, however,
revealed that the simple projectors do not correspond
to D-branes but we need some modification due to the midpoint
singularity.

There has been a proposal which takes exactly the same form
but has a different interpretation.  In \cite{r:KMW}, 
it was proved that the (closed string) boundary state satisfies
\begin{equation}\label{idempotency}
 \Psi\star\Psi = \mathcal{C} \Psi\,
\end{equation}
where $\mathcal{C}$ is a pure ghost term with divergent
coefficient.  
The star product $\star$ here is that for closed string
and there are two vertices studied rather carefully in the
literature \cite{r:nonpol} \cite{r:HIKKO}.
A notable feature of the identity (\ref{idempotency})
is that it holds for both vertices (up to the coefficient $\mathcal{C}$
on the right hand side).
It was argued that this relation is closely connected to
the factorization relation in boundary CFT \cite{r:KM}
and it is mostly independent of the definition of the vertices
and the background as long as the conformal symmetry is maintained.  
It follows that
any sensible boundary state in any consistent background 
needs to satisfy this nonlinear relation.  In this sense, this
relation gives a precise definition of D-brane in SFT language.

A difficulty was the interpretation of the result.
Since this is closed string field theory, it is rather
subtle to consider the tachyon vacuum.  In this sense, it is
rather hard to apply VSFT like interpretation.
Instead of that, combined with the relation for OBS,
we show that it is more natural to see them as
singular limits of the effective open/closed string field theory
which is dual to Witten's OSFT.

The organization of this note consists of two parts.
In the first part, we present the definition
of OBS and some relations among them in section 2. We have to mention
that the discussion given in this proceeding is rather preliminary.
It is clear that the open boundary state is definitely
important apart from the interpretation in SFT and deserves
an independent research.  We will soon give more detailed
analysis in a forthcoming paper \cite{r:IIM}.

In the second part, we focus on the application to VSFT like theory.
In section 3, we will show that the OBSs defined in section 2
satisfy the same relation with respect to the open string
star products.  This is a natural generalization
of the relation (\ref{idempotency}) for the closed string
boundary states. We present the explicit computation for
light-cone type vertex.  We note that a state which coincides with
OBS was proposed in \cite{r:GRSZ}. It is introduced as
a variant of the projector with respect to Witten-type
vertex.  Although our definition of OBS has more freedom
and the interpretation is different, the proof of the
idempotency is basically the same as theirs.
The feature that the projector relation holds for various vertices
distinguishes OBS from other projectors in Witten's OSFT 
such as sliver, butterfly, or identity. 
In section 4, we describe the modular dual description of
Witten's open string field theory and argue that our projector
equations should not be interpreted as the equation for VSFT
like scenario but rather as the singular limit of such an effective
open/closed string field theory.

\section{Construction of open boundary state}
\paragraph{General arguments}
The definition of the open boundary state can be derived from
the closed string sector (\ref{bc2}) by the doubling technique.
Let us take the energy momentum tensor as the first example.
We consider an open  world sheet with rectangular shape
($0\leq \sigma\leq 2\pi$, $\tau\geq 0$). We identify the
anti-holomorphic part of the energy-momentum tensor as,
\begin{equation}
 \bar T(\bar w)=T(2\pi-\bar w)\,, \quad
w=\sigma+i\tau\,.
\end{equation}
The combined tensor field $T(w,\bar w)=T(w)+\bar T(\bar w)$
then satisfies the boundary condition (\ref{bc1}) at the boundary
$\sigma=0,\pi$.  In terms of mode operators, 
it is equivalent to a replacement $\bar L_n\rightarrow L_n$.
The boundary condition at $\tau=0$ is then replaced by a
condition to the state,
\begin{equation}\label{naive}
(T(\sigma)-\tilde{T}(-\sigma))|B^o\rangle=0
\end{equation}
for $0<\sigma<\pi$.
This equation has a subtlety at $\sigma=0,\pi$ and for the
detail see \cite{r:IIM}.

In order to define an open boundary state, we have to specify
three boundary conditions, two D-branes at $\sigma=0,\pi$
and a D-brane at $\tau=0$.  The first two boundary conditions
in general are implemented by the doubling
(for a generic conformal fields $\phi,\bar\phi$),
\begin{equation}
 \bar\phi_\alpha(\bar w)=(\Lambda_r)_{\alpha\beta}\phi_\beta(2\pi-\bar w)\,,
\end{equation}
and the periodicity of the chiral field $\phi_\alpha$ is
\begin{equation}
 \phi_\alpha(w+2\pi)=G_{\alpha\beta}\phi_\beta(w)\,.
\end{equation}
The boundary condition at $\sigma=0$ is then specified by,
\begin{equation}
 \bar\phi_\alpha(\overline{i\tau})=
(\Lambda_{r} G)_{\alpha\beta}\phi_\beta(i\tau)
\equiv (\Lambda_{l})_{\alpha\beta}\phi_\beta(i\tau)\,.
\end{equation}
The boundary condition at $\tau=0$ can be defined similarly as,
\begin{equation}
 \bar \phi_\alpha(\sigma) =
(\Sigma_b)_{\alpha\beta} \phi_\beta(\sigma),\quad
\sigma\in[0,\pi]
\end{equation}
but it is used as a constraint to the boundary states
$|B^o\rangle_{\Lambda_{l}\Lambda_{r}}^{\Sigma_{b}}$ as,
\begin{equation}
\left(\phi_\alpha(2\pi-\sigma)-
(\Lambda^{-1}_r\Sigma_b)_{\alpha\beta}\phi_\beta(\sigma)\right)
|B^o\rangle_{\Lambda_{l}\Lambda_{r}}^{\Sigma_{b}}=0\,.
\end{equation}

\paragraph{Free bosons}
Let us use the free boson system to illustrate these ideas.
For a free boson, possible boundary conditions are Neumann
and Dirichlet. In terms of $J(w)=\partial X(w)$, they
can be written as,
\begin{equation}
 J(w)+\epsilon \bar J(\bar w)= 0\,,\quad
\mbox{at Re}(w)=0\,.
\end{equation}
Here $\epsilon=1$ for the Neumann and $\epsilon=-1$ for the 
Dirichlet boundary condition.  In the notation of 
our general discussion, the reflection matrices
are identified as $\Lambda_{r,l}=-\epsilon$.
Since $G=\epsilon_l\epsilon_r$, the chiral field becomes
periodic when $\epsilon_l=\epsilon_r$ and anti-periodic when
$\epsilon_l=-\epsilon_r$.
It gives the well-known free boson mode expansions,
\begin{eqnarray}
X^{(NN)}(w,\bar w)&=&\hat x-\alpha' \hat p(w-\bar w)
\nonumber\\
&&~~~~~~~
+i\left(\frac{\alpha'}{2}\right)^{1/2} \sum_{m\neq 0}\frac{1}{m}
\alpha_m(e^{imw}+e^{-imw})\,,\\
X^{(DD)}(w,\bar w) &=& 
x+\frac{y-x}{2\pi}(w+\bar w)
\nonumber\\
&&~~~~~~~~+
i\left(\frac{\alpha'}{2}\right)^{1/2} \sum_{m\neq 0}\frac{1}{m}
\alpha_m(e^{imw}-e^{-imw})\,,\\
 X^{(DN)}(w,\bar w) &=&x +i\left(\frac{\alpha'}{2}\right)^2
\sum_{r\in Z+1/2}\frac{1}{r}\alpha_r(e^{irw}-e^{-ir\bar w})\,,\\
 X^{(ND)}(w,\bar w) &=&x +i\left(\frac{\alpha'}{2}\right)^2
\sum_{r\in Z+1/2}\frac{1}{r}\alpha_r(e^{irw}+e^{-ir\bar w})\,.
\end{eqnarray}
The commutation relations for mode variables are,
\begin{eqnarray}
 [\alpha_n,\alpha_m]=n\delta_{n+m,0}\,,\quad
 [\hat x,\hat p]=i\,.
\end{eqnarray}

The condition for the boundary state can be written 
in similar fashion to the closed string case,
\begin{equation}\label{bc3}
 \partial_\tau X(\sigma,\tau)|_{\tau=0}|B^o\rangle
=0\  \mbox{(Neumann)}\,,\qquad
 \partial_\sigma X(\sigma,\tau)|_{\tau=0}
|B^o\rangle=0\ \mbox{(Dirichlet)}\,.
\end{equation}
The only difference is the mode expansion depends on the boundary
conditions for open strings.
We note that there are zero-mode quantum operators
only in NN sector.

The open boundary states are obtained by solving (\ref{bc3}).
The result is written compactly as,
\begin{eqnarray}
|B^o\rangle^{\epsilon_b}_{\epsilon_l\epsilon_r}
&\propto& \exp\left(
-\epsilon_l\epsilon_b\sum_{n>0} \frac{1}{2n}
\alpha_{-n}^2
\right)|\mbox{zero mode}\rangle_{\epsilon_l,\epsilon_r}^{\epsilon_b}\,.
\end{eqnarray}
$\epsilon_{l,r,b}$ describe the boundary conditions
at left, right and bottom respectively and $+1$ for Neumann boundary
condition and $-1$ for Dirichlet boundary condition.
$n$ runs over positive integers when $\epsilon_l=\epsilon_r$
and half-odd positive integers when $\epsilon_r\neq \epsilon_r$.
The vacuum state $|\mbox{zero mode}\rangle_{\epsilon_l,
\epsilon_r}^{\epsilon_b}$ is the simple Fock vacuum when 
$(\epsilon_l,\epsilon_r)\neq (+1,+1)$. For 
$(\epsilon_l,\epsilon_r)= (+1,+1)$, we need to prepare
the zero mode wave function.  An appropriate choice is
$|\hat p=0\rangle$ for $\epsilon_b=1$ and
$|\hat x=x_0\rangle$ for $\epsilon_b=-1$.

A consistency check of these states is to calculate the inner product
between them and compare it with the usual path integral formula.
It gives a disc amplitude where the four sides are specified by 
various boundary conditions. Although this is a disc diagram,
we can expect to have an analogue of the modular invariance.
In particular, it should be written as,
\begin{equation}\label{modular}
 {}_{\Lambda_r,\Lambda_l}^{\Sigma_t}\langle B^o| q^{L_0}
|B^o\rangle_{\Lambda_l,\Lambda_r}^{\Sigma_b}
\propto
 {}_{\Sigma_b,\Sigma_t}^{\Lambda_r}\langle B^o| \tilde q^{L_0}
|B^o\rangle_{\Sigma_t,\Sigma_b}^{\Lambda_l}\,,
\end{equation}
where $q=e^{\pi it}, \tilde q=e^{-\pi i/t}$ for a real $t>0$.

For the free case, we can confirm this relation easily\footnote{
A useful formula is,
$$
\langle 0|e^{q a^2/2}e^{\pm (a^\dagger)^2/2}|0\rangle=
(1-q)^{\mp 1/2}\,
$$
for a simple oscillator  $[a,a^\dagger]=1$.
}.
A straightforward computation gives,
\begin{equation}\label{inner}
 {}^{\epsilon_t}_{\epsilon_r\epsilon_l}\langle B^o 
| q^{L_0}|B^o\rangle^{\epsilon_b}_{\epsilon_l\epsilon_r}
=( \prod_{n>0}(1-\epsilon_t\epsilon_b q^{2n}))^{-1/2}.
\end{equation}
As we have used, $n$ becomes positive integer or half-odd positive 
number depending on $\epsilon_l,\epsilon_r$.
(We need to add zero point energy, zero-mode contribution).
Since these expressions can be written in terms of theta functions,
(\ref{modular}) reduces to the standard modular transformation law.

One interesting aspect of these computations is that
the disc amplitude may have such a modular property.
The complication should come from the conformal anomaly,
namely the mapping from disc to rectangle
through conformal mapping is given by Jacobi's elliptic function.
However, an easier alternative computation comes from
the path integral formula. These analyses will be discussed 
in more detail in \cite{r:IIM}.
We take the  case where all the boundary conditions are Neumann.
On the rectangle, $X$ has the mode expansion,
\begin{equation}\label{me}
 X(\sigma,\tau) = \sum_{m,n\geq 0} X_{mn}\cos(m\sigma)
\cos(n\tau/t)\,.
\end{equation}
We then evaluate the gaussian integration,
\begin{equation}\label{ptf}
 Z=\int dX_{mn}\exp\left(-\frac{1}{4\pi\alpha'}
\int d^2\sigma (\partial X)^2
\right)\,.
\end{equation}
Plugging the expansion (\ref{me}) into (\ref{ptf}) and performing
the gaussian integration over $X_{mn}$, we obtain,
\begin{equation}
 Z\propto \prod_{m,n\geq 1} (m^2 t+n^2/t)^{-1/2}
\propto \eta(q^2)^{-1/2}\,.
\end{equation}
The computation for other boundary conditions is exactly similar
and reproduces (\ref{inner}).

\paragraph{Ghosts}
For the discussion of string field theory in the following sections,
we need the open boundary state for the reparametrization ghosts.
Unlike the Majorana fermions which turn out to be more nontrivial,
the open boundary state for the ghost fields can be straightforwardly
obtained.  The boundary conditions become,
\begin{equation}
 (b_n-b_{-n})|B^o\rangle^{(gh)}= (c_n+c_{-n})|B^o\rangle^{(gh)}=0
\,.
\end{equation}
These conditions can be solved immediately as,
\begin{equation}
 |B^0\rangle^{(gh)}=
\exp\left(\sum_{n>0}c_{-n}b_{-n}\right)
c_0 c_{1}|0\rangle^{(gh)}\,,
\end{equation}
where $|0\rangle^{(gh)}\rangle$ is the $SL(2,R)$-invariant
ghost vacuum.
The inner product between the ghost OBS becomes,
\begin{equation}
 {}^{(gh)}\langle B^o| b_0 q^{L_0}|B^0\rangle^{(gh)}
=\prod_{n=1}^\infty (1-q^n)\,,
\end{equation}
which is again the square-root of the annulus amplitude
from the ghost sector.

\section{Projector equation for OBS}
We first mention a role of OBS in string field theory
context which is different from the study in other parts.
As in the closed string case, it  is natural to use it as
the source term of string field theory.  It can be rewritten in
the form,
\begin{equation}
 Q\Psi+g\Psi\star\Psi= |B^o\rangle\,.
\end{equation}
This equation generalizes the Yang-Mills equation
in the presence of the source D-brane.
We will pass the careful study of this relation in our future
publication and will study a different feature
of boundary states in the context of string field theory.

In the analogy with closed string field theory \cite{r:KMW},
we expect that the idempotency relation holds for OBS.
In this paper, an explicit computation is made only for
light-cone type star product \cite{r:HIKKO}.
Computation for Witten type vertex would be similar
and was made in \cite{r:GRSZ} for a particular type of
OBS.

We first derive the star product of OBSs which have the following
form,
\begin{eqnarray}\label{g-obs}
 |\Phi_r\rangle=
 \exp\Biggl(\frac{1}{2}a^{(r)\dagger}Ma^{(r)\dagger}+\lambda_r a^{(r)\dagger}
 +c^{(r)\dagger}M_gb^{(r)\dagger}\Biggr)
 b^{(r)}_0|x_r^i,p_r^{\mu};\alpha_r\rangle
\end{eqnarray}
where $r=1,2$. Note that we have added the anti-ghost zero mode oscillator
$b_0$ because the physical sector of string fields should belong to
the anti-ghost zero mode sector.
We take the inner product between 3-string vertex and
the tensor product of two bras of OBSs
obtained by imposing the reflectors on the kets.
By carrying this out we arrive at the following formula 
by utilizing several formulae in \cite{r:KMW},
\begin{eqnarray}
 |\Phi_1\star\Phi_2\rangle
 &=&ce^{H_m}\mathcal{C}e^{H_g}
 \delta^{D-d}(x_1-x_2)~b_0|x_1^i,(p_1+p_2)^{\mu};\alpha_1+\alpha_2\rangle\,.
\label{general}
\end{eqnarray}  
Here the notation is,
\begin{eqnarray}
 c&=&\textrm{det}(1-\tilde{M}\tilde{N})^{-\frac{1}{2}}
 \textrm{det}(1-\tilde{N_g}\tilde{M_g}) \nonumber\\
 H_m&=&\frac{1}{2} a^{\dagger} \tilde{M}^{33} a^{\dagger}
 +\tilde{N}^3 \mathbf{P} a^{\dagger}
 -\frac{\tau_0}{2\alpha_1\alpha_2\alpha_3} \mathbf{P}^2 \nonumber\\
 &{}&~~+\frac{1}{2} \tilde{\lambda} \tilde{N}
 (1-\tilde{M}\tilde{N})^{-1} \tilde{\lambda}
 +\frac{1}{2}\mu\tilde{M}(1-\tilde{N}\tilde{M})^{-1}\mu
 +\tilde{\lambda}(1-\tilde{N}\tilde{M})^{-1}\mu\nonumber\\
 H_g&=&c^{\dagger}C^{\frac{1}{2}}\tilde{N}^{33}
 C^{-\frac{1}{2}}b^{\dagger}
 +\sigma(1-\tilde{M}_g\tilde{N}_g)^{-1}\tilde{M}_g\rho\\
 \rho^r&\equiv&\tilde{N}_g^{r3}b^{\dagger},~~
 \sigma^r\equiv c^{\dagger} \tilde{N}_g^{3r}.\nonumber
\end{eqnarray}
In the derivation of this formula, we don't use 
the specific form of $M$ .
In this sense this gives the general formula for the star products of
the squeezed states of the form (\ref{g-obs}).

In a similar fashion as \cite{r:KMW}, we evaluate
$H_m,H_g,\mathcal{C}$. Since our purpose is to evaluate the star products
of OBSs, we set the coefficient matrices $M$ to be $\pm 1$ (unit
matrix) and $M_g$ to be 1. 
We can evaluate these terms using the explicit 
forms of the Neumann functions which are given in \cite{r:KMW}.
The results are : 
\begin{eqnarray}
 &{}&H_m(M=1)=\frac{1}{2}a^{\dagger}a^{\dagger} \label{hm1}\\
 &{}&H_m(M=-1)=-\frac{1}{2}a^{\dagger}a^{\dagger} 
 -\mathbf{P}B^Ta^{\dagger}-\frac{1}{4}\mathbf{P}B^T B\mathbf{P} \label{hm-1}\\
 &{}&H_g(M_g=1)=c^{\dagger}b^{\dagger}\,. \label{hg}
\end{eqnarray}
The computation here is made for the string with NN boundary condition.
In order to calculate the star products of OBSs of different types 
we should construct the SFT for strings with different boundary
conditions from NN. 

In the case of $|B^o\rangle^D_{NN}$, by using (\ref{hm1}),(\ref{hg}) we
obtain
\begin{eqnarray}
 |B^o\rangle^D_{NN}\star|B^o\rangle^D_{NN}
 =c_D|B^o\rangle^D_{NN},
\end{eqnarray}
where $c_D=V_{D-d}c\frac{\partial}{\partial b_0}$
and $V_{D-d}$ is the volume of Dirichlet directions.

In the case of $|B^o\rangle^N_{NN}$, by using (\ref{hm-1}),(\ref{hg}) 
and noting that $\mathbf{P}=0$  we obtain
\begin{eqnarray}
 |B^o\rangle^N_{NN}\star|B^o\rangle^N_{NN}
 =c_N|B^o\rangle^N_{NN},
\end{eqnarray}
where $c_N=c\frac{\partial}{\partial b_0}$.

\section{A modular dual description of Witten's OSFT?}
\begin{figure}[bth]
\centerline{\epsfig{file=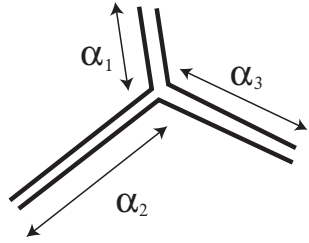,width=40mm}}
\caption{An example of Kaku's vertex}\label{fig:Kaku}
\end{figure}

In the previous section, we have proved the idempotency
of OBS with respect to a specific (light-cone type)
open string vertex. If we combine it with the results of 
\cite{r:KMW, r:KM}, however, it is natural to expect
that the following generalized relations hold for
more generic vertices,
\begin{eqnarray}\label{idemg}
 |B^c\rangle \star |B^c\rangle \sim |B^c\rangle\,,\quad
 |B^o\rangle \star |B^o\rangle \sim |B^o\rangle\,,\quad
 |B^c\rangle \star |B^o\rangle \sim |B^o\rangle\,.
\end{eqnarray}
Here $\star$ products are vertices for closed string, open string,
and open/closed string field theories.
The  basic reason that these relations should hold is
that the boundary states define how left-mover and right-mover
should be attached at every point with specific $\tau$
(world sheet time), namely take the following form,\footnote{
for the closed string channel:  for the open string channel,
as we have seen $X^L$ and $X^R$ are related through
the boundary conditions at $\sigma=0,\pi$.}
\begin{equation}
|B\rangle \sim \prod_\sigma  \delta(X^L(\sigma)-X^R(\sigma))\,.
\end{equation}  
On the other hand, the string vertex
specifies the {\em local} overlap of three (or even more) strings,
which can be written schematically as,
\begin{equation}
 |V_3\rangle = \prod_{a=L,R}
\prod_{i<j}\prod_{\sigma_{j}}\delta(X^a_{i}(\Theta_{ij}(\sigma_j))
-X^a_{j}(\sigma_{j}))\,,
\end{equation}
where $\Theta_{ij}$ is a map from $\sigma$ parameter
of $j$-th string to that of $i$-th string where
they overlap.
By a simple inspection, it is easy to see that (\ref{idemg})
holds generally with singular coefficients for more general 
string vertex considered by M.Kaku \cite{r:Kaku}.
In his proposal, the vertex is not restricted to
the midpoint interaction (as Witten's SFT) and
endpoint interaction (as lightcone SFT) but
overlap can be changed continuously with
extra parameters $\alpha_i$ ($i=1,2,3$) which
describe the length of overlaps as
shown in figure \ref{fig:Kaku}.

A natural question is whether these projector equations
can be interpreted as equations of motion of some 
string field theory.  If this is possible, it gives
the string field theory which is obeyed by
D-branes. This question has been considered
in \cite{r:KMW,r:KM} in the framework of HIKKO type
string field theory and an interpretation similar
to VSFT was given. However, it has not been successful
in reproducing string diagrams.  Especially with only
the boundary states in the closed string channel,
only the vacuum diagrams are generated in
the open string channel. Another drawback 
is that the numerical coefficient of the kinetic term always diverges and the
propagator needs to have a divergent coefficient.

The fact that these identities are satisfied by
more general vertices implies that 
there may be an
interpretation which is different from VSFT.  
Let us recall the most basic
feature of diagrams which are generated by
Witten's open string field theory ---
the existence of the midpoints.  
The perturbation diagrams are essentially fat diagrams of
$\Phi^3$ theory whose center is drawn by the midpoint.

\begin{figure}[b]
\centerline{\epsfig{file=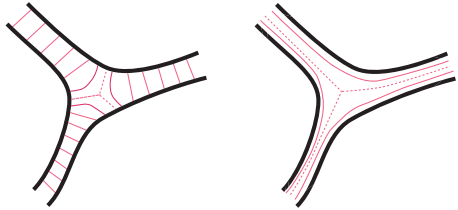,width=60mm}}
\caption{Modular dual picture}\label{fig:dual}
\end{figure}

\begin{figure}[b]
\centerline{\epsfig{file=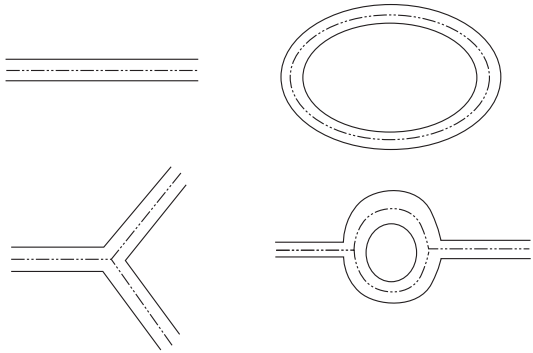,width=70mm}}
\caption{A few diagrams which gives the effective action}\label{fig:dOSFT}
\end{figure}

In order to describe the boundary degree of freedom,
however, it is more natural to see the diagram
in a modular dual fashion.
Namely the time evolution should be traced from the 
{\em boundary} (fig.\ref{fig:dual}).
In such a dual picture, every strip from the
boundary has the same width $\pi/2$ and is attached
at the $\Phi^3$ diagram drawn by the midpoint.
In this dual picture, the external states are the boundary
states with time evolution,
\begin{equation}\label{modif}
 \Phi=e^{-\pi (L_0+\tilde L_0)/2}|B^c\rangle_\alpha,\quad
\mbox{or}\quad e^{-\pi L_0/2}|B^o\rangle_\alpha\,.
\end{equation}
We note that there is an extra parameter $\alpha$ 
in order to describe the length of the closed (open) strings.
For the open string boundary state attached with
the external line, we need to take $\alpha\rightarrow \infty$
limit.  These states can be used to construct string amplitude
by taking the inner product with ``the string vertex''
which describes the attachment of these states at the 
$\Phi^3$  graph drawn by the midpoint,
\begin{equation}
 A\sim {}_{\alpha_1,\cdots,\alpha_N}\langle V_N
|\Phi_1\rangle_{\alpha_1}\cdots|\Phi_N\rangle_{\alpha_N}\,.
\end{equation}
We note that only the planar diagram can be reproduced by
such an inner product.  In this way, we have arrived at
an ``effective action'' which describes the planar diagrams
of Witten's SFT in the dual channel, (fig.\ref{fig:dOSFT})
\begin{equation}
 \Gamma(\Phi)\sim(\Phi^o,\Phi^o)+(\Phi^c,\Phi^c)+(\Phi^o,\Phi^o,\Phi^o)
+(\Phi^o,\Phi^o,\Phi^c)+(\Phi^c,\Phi^c,\Phi^c)+\cdots\,.
\end{equation}
We note a few features of such an effective theory,
\begin{enumerate}
 \item There is no cohomology in the
       quadratic part  $\Phi^2$. In this sense, it is
       similar to VSFT in that it does not generate the usual propagator.
       However, the interpretation is totally
       different.  This is not the expansion from tachyon vacuum
       of the existing open or closed SFT.  Instead of that,
       this is simply a re-interpretation of Witten's open string
       field theory in the dual channel.
 \item The moduli space of string perturbations 
       are generated by the parameters in the vertices
       instead of the propagator.  
 \item We need a nonpolynomial action in order to describe
       all the planar diagrams. This feature resembles
       nonpolynomial closed string field theory developed in
       \cite{r:nonpol}.
 \item On-shell condition for the external states becomes,
\begin{equation}\label{nonpoly}
 \Phi+\Phi^2+\cdots=0
\end{equation}
       which is different from (\ref{idemg}) in the inclusion of
       higher nonlinear terms.  This is interpreted as
       the modification due to (\ref{modif}).  Such modification
       is needed anyway to regularize the coefficients appearing
       in the equation \cite{r:KMW}.
\end{enumerate}
It is of course rather nontrivial to show explicitly
that the modified
boundary states (\ref{modif}) satisfy the nonpolynomial
equation (\ref{nonpoly}).  It is however plausible
since this is just the rewriting of the Feynman diagrams
of Witten's open SFT.

%


\section*{Acknowledgments}

We would like to thank I. Kishimoto for the valuable comments
on this work.  We are also obliged to Y. Imamura for the
collaboration on the open boundary states.

\newcommand{\spi}{{\sqrt{\pi}}}
\newcommand{\refl}{{\langle R(1,2)|}}
\newcommand{\vtx}{{|V(1,2,3)\rangle}}

\section*{Appendix: calculation of the star product}
The light-cone type star product is defined by the reflector $\refl$ which
maps a ket vector to a bra vector and the three string vertex 
$|V(1,2,3)\rangle$ which lives in the tensor products of three open string
Hilbert spaces:
\begin{eqnarray}
 |\Psi_1\star\Psi_2\rangle_3&=&\int db_0^{(1)}db_0^{(2)}
 {}_1\langle\Psi_1|{}_2\langle\Psi_2|V(1,2,3)\rangle,\\
 {}_2\langle\Phi|&=&\int db_0\refl\Phi\rangle_1. \\
\end{eqnarray}
The reflector is defined by \cite{r:HIKKO}
\begin{eqnarray}
 \refl&\equiv&
 \int d^d x^{(1)}d^d x^{(2)}\frac{d\alpha_1d\alpha_2}{(2\pi)^2}
 {}_1\langle x^{(1)},\alpha_1|{}_2\langle x^{(2)},\alpha_2|\nonumber\\
 &{}&~~\exp\Biggl[\sum_{n\geq 1}
 (-)^{n+1}\bigl(a_n^{(1)}a_n^{(2)}
 +c^{(1)}_n b^{(2)}_n-b^{(1)}_n c^{(2)}_n\bigr)\Biggr] \nonumber\\
 &{}&\times\delta^d(x^{(1)}-x^{(2)})
 \delta(b_0^{(1)}-b_0^{(2)})2\pi\delta(\alpha_1+\alpha_2).
\end{eqnarray}
The three string vertex $|V(1,2,3)\rangle$ is given explicitly in terms of
oscillators as: 
\begin{eqnarray}
 \vtx&\equiv&\int~\delta(1,2,3)\mu(1,2,3)
 \prod_{r=1}^{3}\bigl(1+\sum_{n=1}^{\infty}\sum_{s=1}^{3}
 w^{rs}_n\alpha_s c_n^{(s)\dagger}b_0^{(r)}\bigr) \nonumber\\
 &{}&e^{F(1,2,3)}
 |p_1,\alpha_1\rangle_1|p_2,\alpha_2\rangle_2|p_3,\alpha_3\rangle_3, \\
 F(1,2,3)&\equiv&
 \sum_{n,m=1}^{\infty}\sum_{r,s=1}^{3}\tilde{N}^{rs}_{nm}
 \Biggl(\frac{1}{2}a^{(r)\dagger}_n a^{(s)\dagger}_m
 +(\sqrt{n}\alpha_r)c_n^{(r)\dagger}
 (\sqrt{m}\alpha_s)^{-1}b_m^{(s)\dagger}\Biggr) \nonumber\\
 &+&\sum_{n=1}^{\infty}\sum_{r=1}^{3}
 \tilde{N}^r_n a^{(r)\dagger}_n\mathbf{P}
 -\frac{\tau_0}{2\alpha_1\alpha_2\alpha_3}\mathbf{P}^2,\\
 \int \delta(1,2,3)&\equiv&
 \int\frac{d^d p_1}{(2\pi)^d}\frac{d^d p_2}{(2\pi)^d}\frac{d^d p_3}{(2\pi)^d}
 \frac{d\alpha_1}{2\pi}\frac{d\alpha_2}{2\pi}\frac{d\alpha_3}{2\pi} \nonumber\\
 &{}&(2\pi)^d\delta^d(p_1+p_2+p_3)(2\pi)\delta(\alpha_1+\alpha_2+\alpha_3), \\
 \mathbf{P}&\equiv&\alpha_1 p_2-\alpha_2 p_1, 
\end{eqnarray}
\begin{eqnarray}
 w_n^{rs}&\equiv&\Bigl(\chi^{rs}\bar{N}^s_n
 +\frac{1}{\alpha_r}\sum_{m=1}^{n-1}\bar{N}^{ss}_{n-m,m}\Bigr)n, \\
 \chi^{rs}&\equiv&\delta_{r,s}\frac{1}{\alpha_r}(\alpha_{r-1}-\alpha_{r+1})
 +\sum^3_{t=1}\epsilon^{rst}, \\
 \mu(1,2,3)&\equiv&\exp{\Bigl(-\tau_0\sum^3_{r=1}\frac{1}{\alpha_r}\Bigr)}, 
\qquad
 \tau_0\equiv\sum^3_{r=1}\alpha_r\ln|\alpha_r|.
\end{eqnarray}
We consider the tensor product
\begin{eqnarray}
 |\Phi_1\rangle \otimes |\Phi_2\rangle &=&
 \exp\Biggl(\frac{1}{2}a\tilde{M}a+\tilde{\lambda}a
 +c\tilde{M}_g b\Biggr)\cdot\nonumber\\
&&~~~~~~~  \cdot b^{(1)}_0| x_1^i,-p_1^{\mu};-\alpha_1\rangle\otimes
  b^{(2)}_0| x_2^i,-p_2^{\mu};-\alpha_2\rangle\,.
 \end{eqnarray}
The corresponding bra state is obtained by applying the reflector,
\begin{eqnarray}
 \langle\Phi_1| \otimes \langle\Phi_2| =
 \langle x_1^i,-p_1^{\mu};-\alpha_1|b^{(1)}_0 \otimes
 \langle x_2^i,-p_2^{\mu};-\alpha_2|b^{(2)}_0
 \exp\Biggl(\frac{1}{2}a\tilde{M}a+\tilde{\lambda}a
 -c\tilde{M}_g b\Biggr)\,,
 \end{eqnarray}
where
\begin{eqnarray}
 a^{\dagger}&=&
 \Bigl(a^{\mu(1)\dagger}_n,a^{\mu(2)\dagger}_n\Bigr)\,,~~
 \textrm{similar notation for }c^{\dagger} \textrm{ and }b^{\dagger}\nonumber\\
 \tilde{M}_{mn}&=&(-)^{m+n}M_{mn},~~
 (\tilde{M_g})_{mn}=(-)^{m+n}(M_g)_{mn}\\
 \tilde{M}^{rs}_{mn}&=&\left(
 \begin{array}{cc}
 \tilde{M}_{mn}&0\\
 0&\tilde{M}_{mn}
 \end{array}
 \right),\,
 (\tilde{M}_g)^{rs}_{mn}=\left(
 \begin{array}{cc}
 (\tilde{M_g})_{mn}&0\\
 0&(\tilde{M_g})_{mn}
 \end{array}
 \right)
\end{eqnarray}
 We take the inner product between this bra state and the 3-string vertex. 
 For this purpose, it is convenient to rewrite the factor in the exponential in the vertex as 
\begin{eqnarray}
 F(1,2,3) &\equiv& 
 \frac{1}{2}a^{\dagger}\tilde{N}a^{\dagger}+\mu a^{\dagger}
 -\frac{\tau_0}{2\alpha_1\alpha_2\alpha_3}\mathbf{P}^2
 +\frac{1}{2}\sum_{n,m=1}^{\infty}\tilde{N}^{33}_{nm}
 a_n^{(3)\dagger}a_m^{(3)\dagger}
\nonumber\\
 &{}& +\sum^{\infty}_{n=1}\tilde{N}^3_na_n^{(3)\dagger}\mathbf{P}
 +c^{\dagger}\tilde{N}_g b^{\dagger}+c^{\dagger}\tilde{\rho}
 +\tilde{\sigma} b^{\dagger}\nonumber\\
&&
 +c^{(3)\dagger}C^{\frac{1}{2}}\tilde{N}^{33}C^{-\frac{1}{2}}
 b^{(3)\dagger}
\end{eqnarray}
where we introduce some notations
\begin{eqnarray}
 \mu^r&=&\tilde{N}^r\mathcal{P}+\tilde{N}^{r3}a^{(3)\dagger}\nonumber\\
 \tilde{N}_g^{rs}&=&
 \alpha_rC^{\frac{1}{2}}\tilde{N}^{rs}C^{-\frac{1}{2}}\alpha_s^{-1}\nonumber\\
 \tilde{\sigma}^r&=&c^{(3)\dagger}
 \alpha_3 C^{\frac{1}{2}} \tilde{N}^{3r} C^{-\frac{1}{2}} \alpha_r^{-1}
 =c^{(3)\dagger} \tilde{N}_g^{3r} \\
 \tilde{\rho}^r&=&
 \alpha_r C^{\frac{1}{2}} \tilde{N}^{r3} C^{-\frac{1}{2}} \alpha_3^{-1}
 b^{(3)\dagger}+w^{3r}\alpha_r b^{(3)}_0
 =\tilde{N}_g^{r3}b^{(3)\dagger} +w^{3r}\alpha_r b^{(3)}_0 \nonumber
 \end{eqnarray}
 By taking the inner product with the aid 
of the useful formulae \cite{r:KMW}, we can arrive at (\ref{general}).


\end{document}